\documentclass[12pt]{iopart}
\usepackage{iopams}  
\begin{document}

\title[Exactly solvable triatomic models]{Exactly solvable models for triatomic-molecular Bose-Einstein Condensates}
\vspace{1.0cm}
\author{ G. Santos$^{1}$, A. Foerster$^2$, I. Roditi$^1$, \\ 
Z. V. T. Santos$^1$ and A. P. Tonel$^3$}
\vspace{1.0cm}
\address{
{\small $^{1}$CBPF - Centro Brasileiro de Pesquisas F\'{\i}sicas, Rio de Janeiro, RJ - Brazil}\\
{\small $^2$Instituto de F\'{\i}sica da UFRGS, Porto Alegre, RS - Brazil} \\
{\small $^{3}$CCET da Universidade Federal do Pampa/Unipampa, Bag\'e, RS - Brazil}

                 }
\ead{gfilho@cbpf.br}
\vspace{2.0cm}

\begin{abstract}
We construct a family of triatomic models for heteronuclear and homonuclear molecular Bose-Einstein condensates.
We show that these new generalized models are exactly solvable through the algebraic 
Bethe ansatz method and derive their corresponding Bethe ansatz equations and energies.

\end{abstract}

\vspace{2.0cm}


\maketitle

\section{Introduction}

Since the pivotal experimental achievement that led to realizations of Bose-Einstein condensates (BECs), using ultracold 
dilute alkali gases \cite{early,angly}, a great effort has been devoted to the understanding of 
new properties of BEC. The creation of a molecular BEC compound, which has been obtained 
by different techniques \cite{mol}, leads one also to the compelling chemistry of BECs, 
where the atomic constituents may form molecules, for instance,  by Feshbach resonances \cite{feshbach} or 
photoassociation \cite{photo}. These results turned the search for integrable models that could be candidates for 
describing  BEC properties into a very active field of research \cite{jon1,jonjpa,key-3,dukelskyy,Ortiz,Kundu,eric5}.   
In fact, exactly solvable models are expected to provide a significant 
impact in this area, a view that has been promoted in \cite{hertier, batchelor}.

Moreover, the recent experimental evidence for Efimov states in an ultracold cesium gas \cite{EfimovEvidence} 
provided a physical ground for the search of tri-atomic homonuclear molecular BECs. Due to the rapid 
technological developments in the field of ultracold systems, this experiment is just the beginning of the study of 
tri-atomic molecules \cite{Hammer,esry}.

We introduce, in the present paper, a complete family of new solvable models for both heteronuclear and 
homonuclear {\it tri-atomic} molecular BECs obtained through a combination of two Lax operators constructed 
using special realizations of the  $su(2)$ and $su(1,1)$ algebras, as well as a multibosonic representation 
of $sl(2)$, discussed recently in \cite{Tomasz}.

Until now, integrability was shown for those Hamiltonians describing  heteronuclear and homonuclear 
{\it di-atomic} molecular BECs \cite{jonjpa}. In the heteronuclear 
case, two different atoms, labelled $a$ and $b$, can be combined to produce a molecule labelled by $c$.  
The different degrees of freedom in such  models are represented by canonical creation and annihilation 
operators $\{a,\,b,\,c{,...,}\,a^{\dagger},\,b^{\dagger},\,c^{\dagger}\,{,...}\}$ satisfying the usual commutation 
relations $[a,\,a^\dagger]=I$, etc., and commuting among different species. 
The Hamiltonian for the {\it di-atomic} heteronuclear model reads \cite{jonjpa,maj}
\begin{eqnarray}
H&=&U_{aa}N_a^2 + U_{bb}N_b^2 +U_{cc}N_c^2+U_{ab}N_aN_b+U_{ac}N_aN_c+U_{bc}N_bN_c \nonumber \\
 &+& \mu_aN_a +\mu_bN_b+\mu_cN_c + \Omega(a^{\dag}b^{\dag}c +c^{\dag}ba).
\label{ham}
\end{eqnarray}
The parameters $U_{ij}$ describe S-wave scattering,
$\mu_i$ are external potentials and $\Omega$ is the amplitude for interconversion of atoms and molecules.  
In the homonuclear case, two identical atoms labelled $a$ are combined to produce a molecule $c$ \cite{jonjpa}.
The generalization to the triatomic case is not immediate and in particular, for the homonuclear case, a one mode multibosonic realization of $sl(2)$ is essential.

In what follows we will show that it is possible to find other integrable generalized molecular BEC models, 
more specifically, heteronuclear and homonuclear {\it tri-atomic} models. We first introduce two models, one 
with two identical species of atoms and a 
different one and another model where there are three different species of atoms. 
In both cases the three atoms can be combined to form a molecule. We then introduce a model 
consisting of three identical species of atoms which uses a recently defined multibosonic 
realization of the $sl(2)$ algebra \cite{Tomasz}.
We present how these generalized models can be derived from a transfer matrix, thus allowing us to 
show their integrability by the algebraic Bethe ansatz method. We also obtain their corresponding 
energies and Bethe ansatz equations.

\section{Triatomic molecular models}

We are going to introduce in this section three different Hamiltonians describing 
triatomic-molecular BECs, 
two for the hetero-atomic case and one for the homo-atomic case. We are considering the 
coupling parameters real, such that the Hamiltonians are hermitian. The $U$ parameters describe the $S$-wave scattering, 
the $\mu$ parameters are the externals potentials and $\Omega$ is the amplitude for interconversion of atoms and 
molecules. The operators $N_j,\;\;j=a,\;b,\;c,\;d$ are the number operators acting in the Fock space.
 
\subsection{Heteronuclear molecular models}

We can construct two models for hetero-atomic molecular BECs, one with two identical species of atoms and a different one and another model 
where there are three different species of atoms. 

\subsubsection{Model for 2 atoms $a$ and 1 atom $b$} 
$ \, $
$ \, $

\noindent The Hamiltonian that describes the interconversion of a heterogeneous tri-atomic molecule labelled by $c$ with two 
atoms of type $a$ and one atom of type $b$ is given by

\begin{eqnarray}
H &=&U_{aa}N_a^2 + U_{bb}N_b^2 +U_{cc}N_c^2+U_{ab}N_aN_b+U_{ac}N_aN_c+U_{bc}N_bN_c  \nonumber \\&+& \mu_aN_a +\mu_bN_b+\mu_cN_c  +  
\Omega(a^{\dag}a^{\dag}b^{\dag}c +c^{\dag}baa).
\label{ham1a}
\end{eqnarray}

This Hamiltonian (\ref{ham1a}) has two independent conserved quantities,

\begin{eqnarray}
I_{1}&=&N_{a}+2N_{c},\nonumber \\
I_{2}&=&N_{a}-2N_{b},
\label{cq2}
\end{eqnarray}
\noindent where $I_2$ is the imbalance between the number of atoms of type $a$ and $b$.  
The total number of particles $N = N_a + N_b + 3N_c$ 
can be written using these conserved quantities,

\[
N= \frac{3I_{1}-I_{2}}{2},
\]
 and it is also conserved.

Writing the $S$-wave diagonal part of (\ref{ham1a}) as a combination of the conserved quantities (\ref{cq2}) we find

$$
\alpha I_{1}^{2} + \beta I_{2}^{2} + \gamma I_{1}I_{2},$$

\noindent where we have used the following identification for the coupling constants

\begin{equation*}
U_{aa}  =  \alpha+\beta+\gamma, \qquad U_{bb}  =  4\beta, \qquad\qquad U_{cc}  =  4\alpha,
\end{equation*}
\begin{equation*}
U_{ab}  =  -4\beta-2\gamma, \qquad U_{ac}  =  4\alpha+2\gamma, \qquad U_{bc}  =  -4\gamma.
\end{equation*}

Therefore the Hamiltonian (\ref{ham1a}) can also be written as

\begin{eqnarray}
H &=& \alpha I_{1}^{2} + \beta I_{2}^{2} + \gamma I_{1}I_{2} \nonumber \\
  &+& \mu_aN_a + \mu_bN_b + \mu_cN_c +  \Omega(a^{\dag}a^{\dag}b^{\dag}c +c^{\dag}baa).
\label{ham1b}
\end{eqnarray} 

\noindent This form will be used later when we show how the Hamiltonian (\ref{ham1a}) can be derived from a transfer matrix 
to establish integrability 
in the general context of the Yang-Baxter algebra.

\subsubsection{Model for 3 different atoms: $a$,\; $b$ and $c$}
$ \, $
$ \, $

\noindent The Hamiltonian  for a model of heterogeneous tri-atomic molecule labelled by $d$ with one atom of type $a$, one 
atom of type $b$ and one atom of type ${c}$ is given by

\begin{eqnarray}
H &=&U_{aa}N_a^2 + U_{bb}N_b^2 + U_{cc}N_c^2+ U_{dd}N_d^2  \nonumber \\ &+& U_{ab}N_aN_b+U_{ac}N_aN_c+U_{ad}N_aN_d+U_{bc}N_bN_c + 
U_{bd}N_bN_d+U_{cd}N_cN_d  \nonumber \\&+& \mu_aN_a +\mu_bN_b+\mu_cN_c + \mu_dN_d +  \Omega(a^{\dag}b^{\dag}c^{\dag}d + d^{\dag}cba).
\label{ham2a}
\end{eqnarray}
This Hamiltonian (\ref{ham2a}) has three independent conserved quantities,

\begin{eqnarray}
I_{1} & = & N_{a}+N_{d}, \nonumber\\
I_{2} & = & N_{b}+N_{d},\label{cq2b}\\
I_{3} & = & N_{c}+N_{d}.\nonumber
\end{eqnarray}

\noindent The physical quantities representing the different imbalances between the number of atoms of different species $(J_{ab}, J_{ac}, J_{bc})$
and the total number of atoms $N$ can be expressed as a combination of these quantities, for example, 

\begin{eqnarray}
J_{ab} & = & N_{a}-N_{b} = I_{1} - I_{2}, \nonumber\\
J_{ac} & = & N_{a}-N_{c} = I_{1} - I_{3},\nonumber\\
J_{bc} & = & N_{b}-N_{c} = I_{2} - I_{3},\nonumber\\ 
N & = & N_a + N_b + N_c + 3N_d = I_{1}+I_{2} + I_{3}, \nonumber
\end{eqnarray}
and are conserved as well.

We can write the $S$-wave diagonal part of the Hamiltonian (\ref{ham2a}) in terms of a combination of the independent conserved quantities (\ref{cq2b}) as

\begin{eqnarray}
\alpha I_{1}^{2}+\beta I_{2}^{2}+\delta I_{3}^{2}+\gamma I_{1}I_{2}+\rho I_{1}I_{3}+\theta I_{2}I_{3},\end{eqnarray}

\noindent where the following identification has been made for the coupling
constants

$$
U_{aa}  =  \alpha, \qquad U_{bb}  =  \beta, \qquad U_{cc}  =  \delta, \qquad U_{dd}  =  \alpha+\beta+\delta+\gamma+\rho+\theta,
$$
$$
U_{ab}  =  \gamma, \qquad U_{ac}  =  \rho,  
\qquad
U_{bc}  =  \theta, \qquad U_{bd}  =  2\beta+\gamma+\theta,\qquad U_{cd}  =  2\delta+\rho+\theta.
$$

Then the Hamiltonian (\ref{ham2a}) can be also written as
\begin{eqnarray}
 H &=& \alpha I_{1}^{2}+\beta I_{2}^{2}+\delta I_{3}^{2}  + \gamma I_{1}I_{2}+\rho I_{1}I_{3}+\theta I_{2}I_{3} \nonumber \\ 
&+& \mu_aN_a + \mu_bN_b + \mu_cN_c + \mu_dN_d + \Omega(a^{\dag}b^{\dag}c^{\dag}d + d^{\dag}cba).
\label{ham2b}
\end{eqnarray}

\subsection{Homonuclear-molecular model}

In the homogeneous case, we can construct a model describing a triatomic-molecular BEC, where all atoms are identical.

\subsubsection{Model for 3 atoms $a$}
$ \, $
$ \, $

\noindent The Hamiltonian that describes the interconversion of a 
homogeneous triatomic molecule labelled by $b$ with three atoms of type $a$ is given by

\begin{eqnarray}
H &=& U_{aa}N_a^2 + U_{bb}N_b^2 + U_{ab}N_aN_b + \mu_aN_a + \mu_bN_b \nonumber \\ 
&+& \Omega(a^{\dag}a^{\dag}a^{\dag} b \, \alpha_{-}(N_a) + \alpha_{-}(N_a) b^{\dag}aaa ).
\label{ham3a}
\end{eqnarray}
\noindent where $\alpha_{-}(N_a)$ is a function of $N_a$ (see next section for more details) 
that controls the amplitude of 
interconversion $\Omega$. This indicates that the density of atoms $N_a$ 
has some influence in the
generation of a bound-state composed by three identical atoms.

This Hamiltonian (\ref{ham3a}) has one conserved quantity,

\begin{eqnarray}
I &=& N_{a} + 3N_{b},
\label{cq3}
\end{eqnarray}
\noindent the total number of particles $N = N_a + 3N_b$.

\section{Integrability and exact Bethe ansatz solution}

In this section we will discuss the derivation, the integrability and the Bethe ansatz solution of these models. 
We begin with the $su(2)$-invariant
$R$-matrix, depending on the spectral parameter $u$: 

\[R(u)= \left( \begin{array}{cccc}
1 & 0 & 0 & 0\\
0 & b(u) & c(u) & 0\\
0 & c(u) & b(u) & 0\\
0 & 0 & 0 & 1\end{array}\right)\]

\noindent with $b(u)=u/(u+\eta)$ and $c(u)=\eta/(u+\eta)$. Above,
$\eta$ is an arbitrary parameter, to be chosen later. It is easy
to check that $R(u)$ satisfies the Yang-Baxter equation

$$
R_{12}(u-v)R_{13}(u)R_{23}(v)=R_{23}(v)R_{13}(u)R_{12}(u-v). 
$$

\noindent Here $R_{jk}(u)$ denotes the matrix acting non-trivially
on the $j$-th and the $k$-th spaces and as the identity on the remaining
space.

Next we define the Yang-Baxter algebra $T(u)$,

\[T(u)= \left( \begin{array}{cc}
 A(u) & B(u)\\
 C(u) & D(u)\end{array}\right)\]

\noindent subject to the constraint

\begin{equation}
R_{12}(u-v)T_{1}(u)T_{2}(v)=T_{2}(v)T_{1}(u)R_{12}(u-v).\label{eq:3}
\end{equation}

\noindent In what follows we will choose different realizations for the monodromy matrix $\pi(T(u))=L(u)$  
to obtain triatomic models for heteronuclear and homonuclear molecular BECs.
In this construction, the Lax operators $L(u)$  have to satisfy the relation

\begin{equation}
R_{12}(u-v)L_{1}(u)L_{2}(v)=L_{2}(v)L_{1}(u)R_{12}(u-v).
\label{RLL}
\end{equation}
\noindent 
Then, defining the transfer matrix, as usual, through

\begin{equation}
t(u)= tr \pi (T(u)) = \pi(A(u)+D(u)),
\label{eq:12}
\end{equation}
\noindent it follows from (\ref{eq:3}) that the transfer matrix commutes for
different values of the spectral parameters; i. e., $[t(u),t(v)]=0$. Consequently, the 
models derived from this transfer matrix will be integrable. Let us now particularize this construction 
for the hetero and homo-atomic molecular cases.

\subsection{Heteronuclear-molecular models}

We may choose the following realization for the two models of heteroatomic molecular BECs

\begin{equation}
\pi(T(u))= L(u) =
u^{-}GL^{S}(u^{-})L^{K}(u^{+}),\label{eq:4a}
\end{equation}

\noindent with $u^{\pm}=u\pm\omega$, $G = diag(+,-)$, the Lax operator $L^{S}(u)$

\[L^{S}(u)=\frac{1}{u} \left( \begin{array}{cc}
u-\eta S^{z} & -\eta S^{+} \\
-\eta S^{-} & u+\eta S^{z} \end{array}\right)\]

\noindent in terms of the $su(2)$ Lie algebra with generators $S^{z}$
and $S^{\pm}$ subject to the commutation relations

$$
[S^{z},S^{\pm}]=\pm S^{\pm},\;\;\;\;[S^{+},S^{-}]=2S^{z},
$$

\noindent and the Lax operator $L^{K}$

\[L^{K}(u)=\frac{1}{u} \left( \begin{array}{cc}
u+\eta K^{z} & \eta K^{-} \\
-\eta K^{+} & u-\eta K^{z} \end{array}\right),\]

\noindent in terms of the $su(1,1)$ Lie algebra with generators $K^{z}$
and $K^{\pm}$ subject to the commutation relations

\begin{equation}
[K^{z},K^{\pm}]=\pm K^{\pm},\;\;\;\;[K^{+},K^{-}]=-2K^{z}.\label{eq:9a}\end{equation}

\noindent Now, using two different realizations for the $su(2)$ and $su(1,1)$ algebras we 
will show how to construct the hetero-atomic Hamiltonians 
(\ref{ham1a}) and (\ref{ham2a}) from the transfer 
matrix (\ref{eq:12}) and present their exact Bethe ansatz solution.

\subsubsection{Model for 2 atoms $a$ and 1 atom $b$}
$ \, $
$ \, $

\noindent Using the following realizations for the $su(2)$ and $su(1,1)$ algebras
$$
S^{+}=b^{\dagger}c,\;\;\; S^{-}=c^{\dagger}b,\;\;\; S^{z}=\frac{N_{b}-N_{c}}{2},$$

$$
K^{+}=\frac{(a^{\dagger})^{2}}{2},\;\;\; K^{-}=\frac{(a)^{2}}{2},\;\;\; K^{z}=\frac{2N_{a}+1}{4},
$$

\noindent it is straightforward to check that the Hamiltonian (\ref{ham1b}) is related
with the transfer matrix $t(u)$ (\ref{eq:12}) through

$$
H = t(u)-\frac{1}{2}u^{-}\eta+\alpha I_{1}^{2}+\beta I_{2}^{2}+\delta I_{1}I_{2},$$

\noindent where the following identification has been made for the parameters

$$
\mu_{a}=u^{-}\eta,\;\;\;\mu_{c}=-\mu_{b}=u^{+}\eta,\;\;\;\Omega=\frac{\eta^{2}}{2}.$$

We can apply the algebraic Bethe ansatz method, using as the pseudovacuum 
the product state $(|0\rangle = |k \rangle \otimes |\phi \rangle $,
with $|k \rangle$ denoting the lowest weight state of the $su(1,1)$ algebra with weight 
$k$, i.e., $K^z |k \rangle = k |k \rangle $ and $|\phi \rangle$ denoting the 
highest weight state of the $su(2)$ algebra with weight $m_z$),
to find the Bethe ansatz equations (BAE)

\begin{equation}
-\frac{(v_{i}-\omega-\eta m_z) (v_{i}+\omega+\eta k)}{(v_{i}-\omega+\eta m_z)(v_{i}+\omega-\eta k)}=\prod_{i\ne j}^{M}\frac{v_{i}-v_{j}-\eta}{v_{i}-v_{j}+\eta},\;\;\; i,j=1,...,M,
\end{equation}
\noindent and the energies of the Hamiltonian (\ref{ham1a}) (see for example \cite{jonjpa,key-3})

\begin{eqnarray}
E & = & (u^{-}-\eta m_z)(u^{+}+\eta k)\prod_{i=1}^{M}\frac{u-v_{i}+\eta}{u-v_{i}} \nonumber \\
& - &(u^{-}+\eta m_z)(u^{+}-\eta k)\prod_{i=1}^{M}\frac{u-v_{i}-\eta}{u-v_{i}} \nonumber \\
 & - & \frac{1}{2}u^{-}\eta+\alpha I_{1}^{2}+\beta I_{2}^{2}+\gamma I_{1}I_{2}.\end{eqnarray}

\subsubsection{Model for 3 different atoms: $a$,\; $b$ and $c$}
$ \, $
$ \, $

\noindent Using the following realizations for the $su(2)$ and $su(1,1)$ algebras

$$
S^{+}=c^{\dagger}d,\;\;\; S^{-}=d^{\dagger}c,\;\;\; S^{z}=\frac{N_{c}-N_{d}}{2},$$

$$
K^{+}=\frac{(a^{\dagger}b^{\dagger})}{2},\;\;\; K^{-}=\frac{(ab)}{2},\;\;\; K^{z}=\frac{N_{a} + N_b + 1}{2},
$$

\noindent it is straightforward to check that the Hamiltonian (\ref{ham2b}) is related
with the transfer matrix $t(u)$ (\ref{eq:12}) through

$$
H(u)  =  t(u)-u^{-}\eta+\alpha I_{1}^{2}+\beta I_{2}^{2}+\delta I_{3}^{2}+\gamma I_{1}I_{2}+\rho I_{1}I_{3}+\theta I_{2}I_{3},
$$

\noindent with the following identification for the parameters

\[
\mu_{a}=\mu_{b}=u^{-}\eta,\;\;\;\mu_{c}=-\mu_{d}=-u^{+}\eta,\;\;\;\Omega=\eta^{2}.\]

Applying the algebraic Bethe ansatz method we find the Bethe ansatz equations (BAE)

\begin{equation}
-\frac{(v_{i} - \omega - \eta m_z)(v_{i} + \omega + \eta k)}{(v_{i} - \omega + \eta m_z)(v_{i} + \omega - \eta k)}=\prod_{i\ne j}^{M}\frac{v_{i}-v_{j}-\eta}{v_{i}-v_{j}+\eta},\;\;\; i,j=1,...,M,
\end{equation}

\noindent and the energies of the Hamiltonian (\ref{ham2a})

\begin{eqnarray}
E & = & (u^{-} - \eta m_z)(u^{+} + \eta k)\prod_{i=1}^{M}\frac{u-v_{i}+\eta}{u-v_{i}} \nonumber \\ &-&(u^{-}  + \eta m_z)(u^{+} - \eta k)\prod_{i=1}^{M}\frac{u-v_{i}-\eta}{u-v_{i}} \nonumber \\
 & - & u^{-}\eta+\alpha I_{1}^{2}+\beta I_{2}^{2}+\delta I_{3}^{2}+\gamma I_{1}I_{2}+\rho I_{1}I_{3}+\theta I_{2}I_{3}.\end{eqnarray}

\subsection{Homonuclear-molecular model}

In this case we may choose the following realization for the Yang-Baxter algebra

\begin{equation}
\pi(T(u))= L(u) =
\eta^{-1}GL^{b}(u - \delta - \eta^{-1})L^{A}(u + \omega),\label{eq:4}
\end{equation}

\noindent with, $G = diag(-,+)$, the Lax operator $L^{b}(u)$

\[L^{b}(u)= \left( \begin{array}{cc}
u + \eta N_{b} & b \\
b^{\dagger} & \eta^{-1} \end{array}\right)\]

\noindent in terms of the canonical boson operators $b$ and $b^{\dagger}$ with $N_b = b^{\dagger}b$, 
subject to the commutation relations

$$
[b,b]=[b^{\dagger},b^{\dagger}]=0,\;\;\;\;[b,b^{\dagger}]=I,
$$
\noindent and the Lax operator $L^{A}$

\[L^{A}(u) = \left( \begin{array}{cc}
u+\frac{\eta}{2} A_{0} & \eta A_{-} \\
-\eta A_{+} & u-\frac{\eta}{2} A_{0} \end{array}\right),\]

\noindent in terms of the $sl(2)$ Lie algebra with generators $A_{0}$
and $A_{\pm}$, subject to the commutation relations

\begin{equation}
[A_{-},A_{+}]= A_{0},\;\;\;\;[A_{0},A_{\pm}] = \pm 2A_{\pm}.\label{eq:9}\end{equation}

We find from the $L$ operator (\ref{eq:4}) the following transfer matrix

\begin{eqnarray}
t(u) &=& -\eta^{-1}(u + \omega + \frac{\eta}{2} A_{0})(u-\delta-\eta^{-1}+\eta N_{b}) \nonumber \\
&+& \eta^{-2}(u + \omega - \frac{\eta}{2} A_{0})+bA_{+}+b^{\dagger}A_{-},
\label{transfer1}
\end{eqnarray}
\noindent with 
\begin{equation}
t(0)=\eta^{-1}\omega(\delta+2\eta^{-1})-\omega N_{b} + \frac{1}{2}(\delta-\eta N_{b})A_{0} +bA_{+}+b^{\dagger}A_{-}.
\label{transfer0}
\end{equation}

There is a one-mode realization of the $sl(2)$ algebra \cite{Tomasz}

$$ A_{0}=\alpha_{0}(N), \;\;\; A_{-} = \alpha_{-}(N)a^l, \;\;\; A_{+} = (a^{\dagger})^l\alpha_{-}(N) $$
\noindent with
\begin{eqnarray}
 \alpha_{0}(N) &=& \frac{2}{l}(N - R) + \alpha_{0}(R) \\
 \alpha_{-}(N) &=& \sqrt{\frac{N!}{(N + l)!}(\frac{1}{l}(N - R) + \alpha_{0}(R))(\frac{1}{l}(N - R) + 1)} 
\end{eqnarray}
\noindent where $N = a^{\dagger}a$ and $l \in \mathbb{N}$. The operator $R$ is

\begin{equation*}
 R = 
\cases{ 0, \qquad\qquad\qquad\qquad\qquad\quad if \qquad l=1 \\ 
\frac{l-1}{2} + \sum_{m=1}^{l-1}\frac{e^{-(2\pi m/l)N}}{e^{(2\pi m/l)} - 1}, \qquad  if \qquad  l>1}
\end{equation*}
\noindent  and acts on the states $\{|n\rangle\}$ as  $R|n\rangle=n\;mod\;l|n\rangle$. The 
function  $\alpha_{0}(R)$ is a function of the spectrum of $R$. For $n=r<l$, we have 

\begin{equation}
 \frac{1}{l}(N - R)|r\rangle = 0|r\rangle
\end{equation}
\noindent with $A_{0} = \alpha_{0}(R)$ such that $\alpha_{0}(R)|r\rangle=\alpha_{0}(r)|r\rangle$ 
and $R|r\rangle=r|r\rangle$. The operator $R$ commutes with $a^l$ and $(a^{\dagger})^l$ and so 
is  a conserved quantity for all models presented in this paper.

Now we will apply this realization to show how to construct the Hamiltonian (\ref{ham3a})  from the transfer 
matrix (\ref{transfer1}) and present their exact Bethe ansatz solution. For $l=3$, which is the case of interest here, it is straightforward to check that the Hamiltonian (\ref{ham3a}) is related with the transfer matrix $t(0)$ (\ref{transfer0}) through 

\begin{equation}
 H = t(0)
\end{equation}
\noindent where we have the following identification

\begin{eqnarray*}
 \eta &=& 9U_{a}+U_{b}-3U_{ab}
\end{eqnarray*}

\begin{eqnarray*}
 6(\omega+\delta) &=& (18U_{a}-2U_{b})N-\eta\rho+18\mu_{a}-6\mu_{b}
\end{eqnarray*}

\begin{eqnarray*}
36\eta^{-1}\omega\delta+72\omega\eta^{-2} &=& \eta\rho^{2}+[(4U_{b}-6U_{ab})N+6(\mu_{b}-3\mu_{a}+\omega)]\rho \nonumber \\ 
&+&4U_{b}N^{2}+12(\mu_{b}+\omega)N 
\end{eqnarray*}
\noindent with $\rho\equiv\rho(R) = 3\alpha_{0}(R)-2R$.
Let $|0\rangle_b$ denotes the Fock vacuum state and let $|r\rangle_A$ denotes the lowest weight state of 
the $sl(2)$ algebra where $r = 0,\;1,\;2,$ are the eingenvalues of $R$ for $l=3$ and $N=nl+r$, with $n \in N$. 
On the product state $|\Psi\rangle = |r\rangle_A \otimes |0\rangle_b$ we can apply the algebraic Bethe ansatz method 
to find the Bethe ansatz equations (BAE)

\begin{equation}
\frac{(1-\eta v_{i}+\eta\delta)(v_{i}+\omega+\frac{\eta}{2}\alpha_{0}(r))}{v_{i}+\omega-\frac{\eta}{2}\alpha_{0}(r)}=\prod_{i\ne j}^{M}\frac{v_{i}-v_{j}-\eta}{v_{i}-v_{j}+\eta},\;\;\; i,j=1,...,M,
\end{equation}

\noindent and the energies of the Hamiltonian (\ref{ham1a}) 

\begin{eqnarray}
E & = & \eta^{-1}(\delta+\eta^{-1})\left(\omega+\frac{\eta}{2}\alpha_{0}(r)\right)\prod_{i=1}^{M}\frac{v_{i}-\eta}{v_{i}} \nonumber \\ &+& \eta^{-2}\left(\omega-\frac{\eta}{2}\alpha_{0}(r)\right)\prod_{i=1}^{M}\frac{v_{i}+\eta}{v_{i}}.\end{eqnarray}

\section{Summary}
We have introduced three new integrable models for both, homogeneous and heterogeneous tri-atomic molecular BECs
obtained through a combination of Lax operators constructed using special realizations of the  
$su(2)$ and $su(1,1)$ algebras and a particular one-mode multibosonic representation of $sl(2)$,
possibilities that were overlooked in previous studies.
The models were solved by means of the algebraic Bethe ansatz method and their corresponding
energies and Bethe ansatz equations were derived.

\subsection*{Acknowledgments}
A. F. thanks J. Links for useful discussions. 
I. Roditi acknowledges FAPERJ
(Funda\c{c}\~ao de Amparo \`a Pesquisa do Estado do Rio de Janeiro).
The authors also acknowledge support from CNPq (Conselho Nacional de Desenvolvimento
Cient\'{\i}fico e Tecnol\'{o}gico).


\section*{References}
\begin{thebibliography}{99}

\bibitem{early} E. A. Cornell and C. E. Wieman, 
Rev. Mod. Phys. \textbf{74} (2002) 875.

\bibitem{angly} J. R. Anglin and W. Ketterle, 
Nature \textbf{416}, (2002) 211.

\bibitem{mol} P. Zoller, 
Nature \textbf{417} (2002) 493.

\bibitem{feshbach} S. B. Papp and C. E. Wieman, cond-mat/0607667.

\bibitem{photo} B. Damski, L. Santos, E. Tiemann, M. Lewenstein, S. Kotochigova, \\ P. Julienne, and P. Zoller, 
Phys. Rev. Lett. {\bf 90} (2003) 110401.

\bibitem{jon1} H.-Q. Zhou, J. Links, M. Gould and R. McKenzie,
J. Math. Phys. \textbf{44} (2003) 4690.

\bibitem{jonjpa} 
J. Links, H.-Q. Zhou, R. H. McKenzie and M. D. Gould, 
J. Phys. \textbf{A36} (2003) R63;

\bibitem{key-3} A. Foerster, J. Links, H.-Q. Zhou, in Classical
and quantum nonlinear integrable systems: theory and applications,
edited by A. Kundu (IOP Publishing, Bristol and Philadelphia,
2003) pp. 208-233.

\bibitem{dukelskyy} J. Dukelsky, G. Dussel, C. Esebbag and S. Pittel,
Phys. Rev. Lett. \textbf{93} (2004) 050403

\bibitem{Ortiz} G. Ortiz, R. Somma, J. Dukelsky and S. Rombouls, 
 Nuclear Physics \textbf{B707} (2005) 421.

\bibitem{Kundu} A. Kundu,
Theoretical and Mathematical Physics \textbf{151} (2007) 831.

\bibitem{eric5} A. Foerster and E. Ragoucy,
Nuclear Physics \textbf{B777} (2007) 373.

\bibitem{hertier} M. H\'eritier,
Nature \textbf{414} (2001) 31.

\bibitem{batchelor} M. T. Batchelor, 
Physics Today \textbf{60} (2007) 36.

\bibitem{EfimovEvidence} T. Kraemer, M. Mark, P. Waldburger, J. G. Danzl, C. Chin, B. Engeser, 
A. D. Lange, K. Pilch, A. Jaakkola, H.-C. Nagerl, R. Grimm,
Nature \textbf{440},(2006) 315.

\bibitem{Hammer} E. Braaten, H.-W. Hammer, Annals of Physics \textbf{322} (2007) 120.

\bibitem{esry} B. D. Esry and C. Greene, Nature \textbf{440} (2006) 289.

\bibitem{Tomasz} T. Goli\'{n}ski, M. Horowski, A. Odzijewicz, A. Sli\.{z}ewska, Jour. Math. Phys. \textbf{48} (2007)  023508.

\bibitem{maj} M. Duncan, A. Foerster, J. Links, E. Mattei, N. Oelkers and A. Tonel, 
Nuclear Physics \textbf{B767} [FS] 227 (2007).


\end{thebibliography}
\end{document}